\documentclass[final,12pt,times,twocolumn]{elsarticle}
\usepackage[margin=1.8cm]{geometry}
\usepackage{amsmath}
\usepackage{amsfonts}
\usepackage{amssymb}
\usepackage{graphicx}
\usepackage{lmodern}
\usepackage{hyperref}
\usepackage{dcolumn}
\usepackage{bm}
\usepackage[mathlines]{lineno}
\usepackage{float}
\usepackage[T1]{fontenc}
\usepackage{subfigure}

\journal{Physics Letters B}

\usepackage{numcompress}

\DeclareUnicodeCharacter{2212}{\ensuremath{-}}

\begin{document}
\newcolumntype{P}[1]{>{\centering\arraybackslash}p{#1}}
\title{\textbf{The $\mu$-$\tau$ mixed symmetry and neutrino mass matrix}}



\author[mymainaddress]{Manash Dey}
\ead{manashdey@gauhati.ac.in}

\author[mymainaddress]{Pralay Chakraborty}
\ead{pralay@gauhati.ac.in}

\author[mymainaddress]{Subhankar Roy\corref{mycorrespondingauthor}}
\cortext[mycorrespondingauthor]{Corresponding author}
\ead{subhankar@gauhati.ac.in}

\address[mymainaddress]{Department of Physics, Gauhati University, India}

\begin{abstract}

We propose an elegant neutrino mass matrix texture entitled \emph{$\mu$-$\tau$ mixed symmetry} highlighting two simple correlations among its elements and a detailed analysis is carried out to see its phenomenological implications. The proposed texture is motivated in the framework of Seesaw mechanism in association with $A_4$ symmetry.
\end{abstract}
\maketitle

The neutrino oscillation is a very peculiar phenomenon in Particle Physics where a neutrino changes its flavour from one to another while it travels from the source to detector. A theory based on quantum mechanics to explain this phenomenon was first proposed by Bruno Pontecorvo in the year 1957 \,\cite{Pontecorvo:1957cp}. In recent past, the SNO\,\cite{SNO:2002tuh}, KamLAND\,\cite{KamLAND:2002uet} and Super-Kamiokande\,\cite{Super-Kamiokande:1998uiq} experiments witnessed the neutrino oscillation. The standard model (SM) of particle physics denies the existence of neutrino mass. The neutrino oscillation, on the other hand strengthens the fact that the three flavour neutrino states\,($\nu_{l=e,\mu,\tau}$) are admixture of three neutrino mass eigenstates\,($\nu_{i=1,2,3}$) having mass eigenvalues\,($m_{i=1,2,3}$). The neutrino oscillation experiments aim to measure the six observational parameters namely, solar mixing angle\,($\theta_{12}$), reactor mixing angle\,($\theta_{13}$), atmospheric mixing angle\,($\theta_{23}$), two mass squared differences\, $\Delta\,m^2_{21}=m_2^2-m_1^2$ and $\Delta\,m^2_{31}=|m_3^2-m_1^2|$ and a Dirac CP violating phase\,($\delta$). Beyond these, there are other two observational parameters termed as Majorana phases\,($\alpha$ and $\beta$) on which the oscillation experiments stay silent. However, it is to be underlined that the two Majorana phases, octant of $\theta_{23}$, neutrino mass ordering and the precise range of $\delta$ are still undetermined in the experiments. 

The extension of SM is manifested in terms of the neutrino mass matrix\,($M_{\nu}$) that appears in the Yukawa Lagrangian\,($\mathcal{L}_Y$). The neutrino mass matrix carries the information of all the physical parameters mentioned above and the theorists in general, try to predict a promising neutrino mass matrix starting from the first principle. Several phenomenological ideas such as $\mu-\tau$ symmetry\,\cite{Harrison:2002er, King:2018kka, Lin:2009bw}, texture zeroes\,\cite{Ludl:2014axa, Borgohain:2020csn}, hybrid textures\,\cite{Kalita:2015tda, Singh:2018bap}, vanishing minors\,\cite{Lashin:2009yd, Araki:2012ip} etc. are attributed to the neutrino mass matrix and are studied extensively from both \emph{bottom-up} and \emph{top-down} viewpoints. This is to be noted that all these approaches insist on reducing the number of independent parameters in the mass matrix either by invoking correlations among the mass matrix elements or restricting the mass matrix elements to take specific values so that the mass matrix becomes predictive. We declare that our work is based on the assumption that neutrinos are Majorana particles\,\cite{Bilenky:2020wjn, EXO-200:2014ofj} and this assumption, in turn forces $M_{\nu}$ to take a complex symmetric texture.

The neutrino physics phenomenology deals fundamentally with two quantities: $M_{\nu}$ and $V$, the Pontecorvo-Maki-Nakagawa-Sakata matrix\,(PMNS matrix)\cite{Maki:1962mu}. We consider two basis systems, namely: \emph{flavour basis}: where the charged lepton mass matrix, $M_{l}$ is diagonal and the \emph{symmetry basis}, where it is not. In general, the PMNS matrix is defined in the flavour basis. If, $M_{l}$ is diagonal, then $V$ is the matrix that diagonalizes $M_{\nu}$: $V^T.M_{\nu}.V= diag(m_1,m_2,m_3)$. If not, then we redefine $V$ as $V=U_{l_L}^{\dagger}.U_{\nu}$, where $U_{l_L}$ is the left-handed charged lepton diagonalizing matrix that diagonalizes $M_{l} M_{l}^{\dagger}$: $U_{l_L}^{\dagger}. M_{l}M_{l}^{\dagger}.U_{l_L}= diag(m_e^2,m_{\mu}^2,m_{\tau}^2)$, and $U_{\nu}$ is the matrix that diagonalizes $M_{\nu s}$: $U_{\nu}^T.M_{\nu s}.U_{\nu}= diag(m_1,m_2,m_3)$. $M_{\nu s}$ is the neutrino mass matrix in the symmetry basis. 

The PMNS matrix $V$, is a $3\times 3$ unitary matrix  and hence contains nine free parameters. As per the PDG parametrization\cite{ParticleDataGroup:2020ssz}, the $V$ is presented as shown below.

\begin{equation}
\label{pmns}
V= P_{\phi}. U. P_M ,
\end{equation}

where, $P_{\phi} =diag (e^{i\phi_1},e^{i\phi_2},e^{i\phi_3})$ and $P_M=diag(e^{i\alpha},e^{i\beta},1)$. The matrix $U$ is represented as shown in the following,

\begin{eqnarray}
U &=& \begin{bmatrix}
1 & 0 & 0\\
0 & c_{23} & s_{23}\\
0 & - s_{23} & c_{23}
\end{bmatrix}\times \begin{bmatrix}
c_{13} & 0 & s_{13}\,e^{-i\delta}\\
0 & 1 & 0\\
-s_{13} e^{i\delta} & 0 & c_{13}
\end{bmatrix}\nonumber\\
&& \quad\quad\times\begin{bmatrix}
c_{12} & s_{12} & 0\\
-s_{12} & c_{12} & 0\\
0 & 0 & 1
\end{bmatrix},
\end{eqnarray}
where, $s_{ij}=\sin\theta_{ij}$ and $c_{ij}=\cos\theta_{ij}$. 
In principle, the three arbitrary phases\,($\phi_{i=1,2,3}$) can be removed from $V$ by redefining the charged lepton fields in terms of these phases. But, this is to be underlined that the redefinition of the charged lepton fields is a choice. In principle, $M_{\nu}$ being a complex symmetric $3\times 3$ matrix, must contain twelve parameters in general. So, it is not unlikely that $M_{\nu}$ may shelter the three unphysical phases $\phi_{1,2,3}$. Here, we highlight that once $V$ is derived from $M_{\nu}$, first it has to be put in the standard form as mentioned in Eq.\,(\ref{pmns}) and thus the unphysical phases $\phi_{1,2,3}$ can easily be isolated so that we can extract the information of mixing angles as well as the other physical phases. But so long the $M_{\nu}$ is concerned, the information of unphysical phases is embedded within it. To simplify, we demarcate $M_{\nu}$ which is the general one, from the $\tilde{M}_{\nu}$ that excludes the unphysical phases. These two matrices are related to each other through the following transformation,

\begin{equation}
\label{gnm}
M_{\nu}=P_{\phi}^*. \tilde{M}_{\nu}. P_{\phi}^*.
\end{equation}
Hence, if we impose certain constraints over the elements of $M_{\nu}$, one cannot deny the involvement of the unphysical phases.

When constraints are put on the neutrino mass matrix, we expect the reduction in the number of free parameters. But at this stage it would have been misleading if we don't mention about whether constraints are imposed on $M_{\nu}$ or $\tilde{M}_{\nu}$. Because for the former, twelve parameters are involved while there are nine parameters for the latter. Hence, depending on the fact whether we are dealing with $M_{\nu}$ or $\tilde{M}_{\nu}$, the scenarios could be different. Because, for the former case, the involvement of unphysical phases is undeniable.  

To understand the significance\,\cite{Xing:2022uax} of the unphysical phases while defining a particular texture, we take the following two examples. First, we take the $\mu$-$\tau$ reflection symmetry\,\cite{Xing:2015fdg, Xing:2020ijf, Xing:2022uax, Liu:2017frs, Duarah:2020zjo, Ganguly:2021swa, Chakraborty:2019rjc} which says,

\begin{eqnarray}
M=\begin{bmatrix}
A & B & B^*\\
B & C & D  \\
B^* & D & C^*
\end{bmatrix},
\end{eqnarray}
where, except $A$ and $D$, all other elements are complex quantities. If the said symmetry is true for $\tilde{M_{\nu}}$, then following relations must hold good,

\begin{eqnarray}
Arg\,[(\tilde{M}_{\nu})_{13}]- Arg\,[(\tilde{M}_{\nu})_{12}]&=&0,\\
Arg\,[(\tilde{M}_{\nu})_{22}]- Arg\,[(\tilde{M}_{\nu})_{33}]&=& 0,\\
Arg\,[(\tilde{M}_{\nu})_{11}]&=& 0,\\
Arg\,[(\tilde{M}_{\nu})_{23}]&=& 0,
\end{eqnarray}

whereas, the same symmetry when imposed on $M_{\nu}$, the following two relations appear,

\begin{eqnarray}
Arg\,[(\tilde{M})_{23}]&=& Arg\,[(\tilde{M})_{12}]+ Arg\,[(\tilde{M})_{13}]\nonumber\\
&&-Arg\,[(\tilde{M})_{11}],\nonumber\\
&=& \frac{1}{2}\,[Arg\,[(\tilde{M})_{22}]+Arg\,[(\tilde{M})_{33}].\nonumber\\
\end{eqnarray}

To extend the discussion on the topic of unphysical phases, we take the following example. A new idea to break the $\mu$-$\tau$ symmetry of the neutrino mass matrix based on phase elements is off late put forward in ref.\,\cite{Chamoun:2019pbh}. The texture of the phase broken $\mu$-$\tau$ symmetry is presented below.

\begin{eqnarray}
\label{nc}
M = \begin{bmatrix}
A & B & B e^{i\psi} \\
B & C & D \\
B e^{i\psi} & D & C e^{i\chi}
\end{bmatrix}.
\end{eqnarray} 
Here, $A$, $B$, $C$ and $D$ are proposed as complex elements. Now the question arises whether $M$ is equivalent to $M_{\nu}$ or $\tilde{M}_{\nu}$. At the first glance, it appears that it is $\tilde{M}_{\nu}$. But on scanning, the number of free parameters in $M$ comes out to be ten (which is more than nine). So we emphasize that $M$ in Eq.\,(\ref{nc}) must contain unphysical phases. Starting from the expression of the general neutrino mass matrix, $M_{\nu}$ mentioned in Eq.\,(\ref{gnm}), we can obtain the texture of $M$ in Eq.\,(\ref{nc}) if following conditions are satisfied.
\begin{eqnarray}
\psi &=& Arg\,[\tilde{M}_{13}]-Arg\,[\tilde{M}_{12}]-\phi_3 +\phi_2,\\
\chi &=& \frac{1}{2}\left( Arg\,[\tilde{M}_{33}]-Arg\,[\tilde{M}_{22]}\right)-\phi_3 +\phi_2\nonumber\\
\end{eqnarray}
Needless to mention that the following two relations,
\begin{eqnarray}
|(\tilde{M})_{12}|&=& |(\tilde{M})_{13}|,\\
|(\tilde{M})_{22}|&=& |(\tilde{M})_{33}|,
\end{eqnarray}
hold good for the two examples discussed above. In our further discussion, we shall stick to $M_{\nu}$.

The general $\mu$-$\tau$ symmetry emphasizes on the correlations: $(M_{\nu})_{13}=(M_{\nu})_{12}$ and $(M_{\nu})_{33}=(M_{\nu})_{22}$. However, these correlations are partially successful in the sense, these are inconsistent with the fact that $\theta_{13}\neq 0$. Twisting the said correlations a little in the following way,  $(M_{\nu})_{13}= -(M_{\nu})_{12}$ and $(M_{\nu})_{33}=(M_{\nu})_{22}^*$, we posit a new texture of the neutrino mass matrix in the flavour basis as follows,

\begin{equation}
 M_{\nu} =
 \begin{bmatrix}
 w & -x & x\\
 -x & y & z\\
 x & z & y^{*}\\
 \end{bmatrix}.
 \label{Mass matrix 1}
\end{equation}

Here, the elements $x$, $y$, $z$ and $w$ are in general complex quantities. We name the above texture as \emph{$\mu$-$\tau$ mixed symmetry} to discriminate it from the  $\mu-\tau$ symmetry\,\cite{Xing:2020ijf, Lin:2009bw, Ma:2005qf}, $\mu-\tau$ reflection symmetry\,\cite{Xing:2015fdg, Xing:2020ijf, Xing:2022uax, Liu:2017frs, Duarah:2020zjo, Ganguly:2021swa, Chakraborty:2019rjc} and $\mu-\tau$ antisymmetry\,\cite{Xing:2015fdg, Xing:2020ijf}. We highlight the following two relations,

\begin{eqnarray}
\phi_2 &=& \frac{1}{4}\left( Arg[(\tilde{M_{\nu}})_{22}] + Arg[(\tilde{M_{\nu}})_{33}]\right)\nonumber\\
&&- \frac{1}{2}\left( Arg[(\tilde{M_{\nu}})_{13}]-Arg[(\tilde{M_{\nu}})_{12}] \right)-\frac{\pi}{2},\nonumber\\
\\
\phi_3 &=&\frac{1}{4}\left(Arg[(\tilde{M_{\nu}})_{22}] + Arg[(\tilde{M_{\nu}})_{33}]\right) \nonumber\\&& + \frac{1}{2}\left(Arg[(\tilde{M_{\nu}})_{13}]- Arg[(\tilde{M_{\nu}})_{12}] \right)+\frac{\pi}{2},\nonumber\\
\end{eqnarray}

which are true with respect to \emph{$\mu$-$\tau$ mixed symmetry}. First, we shall try to estimate the faithfulness of the proposed texture in the light of experiment and later, shall try to understand the first principle associated with this symmetry.   
	
\begin{figure*}
  \centering
    \subfigure[]{\includegraphics[width=0.317\textwidth]{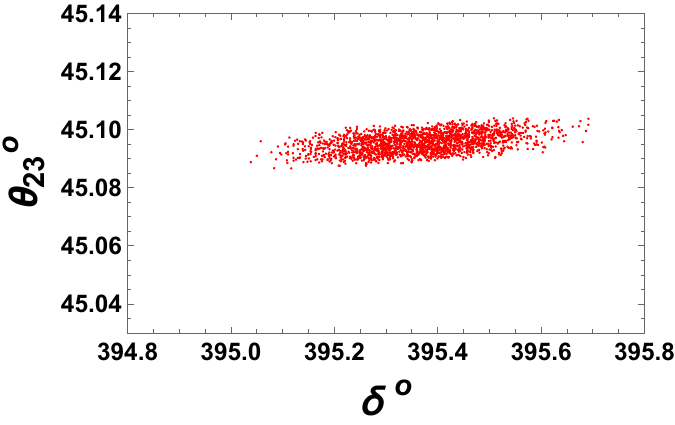}\label{fig:1(a)}} 
    \subfigure[]{\includegraphics[width=0.30\textwidth]{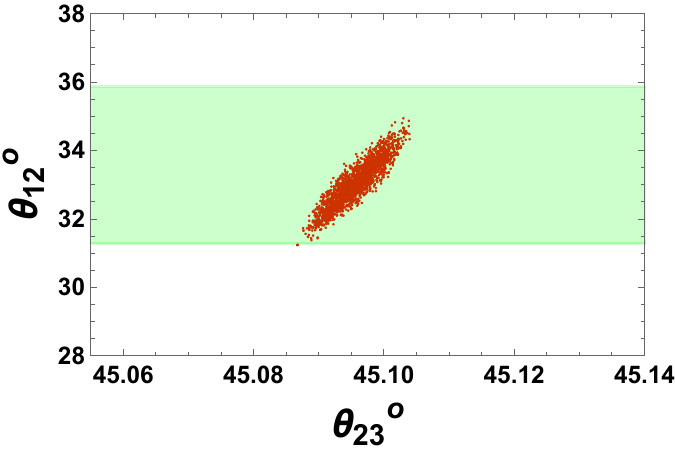}\label{fig:1(b)}} 
    \subfigure[]{\includegraphics[width=0.30\textwidth]{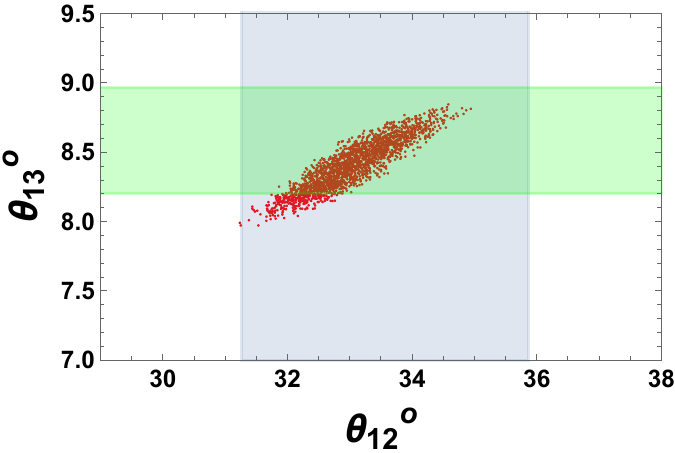}\label{fig:1(c)}}
    \caption{The correlation plots between (a) $\theta_{23}$ and $\delta$. (b) $\theta_{12}$ and $\theta_{23}$. (c)  $\theta_{13}$ and $\theta_{12}$ for normal ordering of neutrino masses with $m_3 = 0.06 \,eV$.}
\label{fig:1}
\end{figure*}

\begin{figure*}
  \centering
    \subfigure[]{\includegraphics[width=0.317\textwidth]{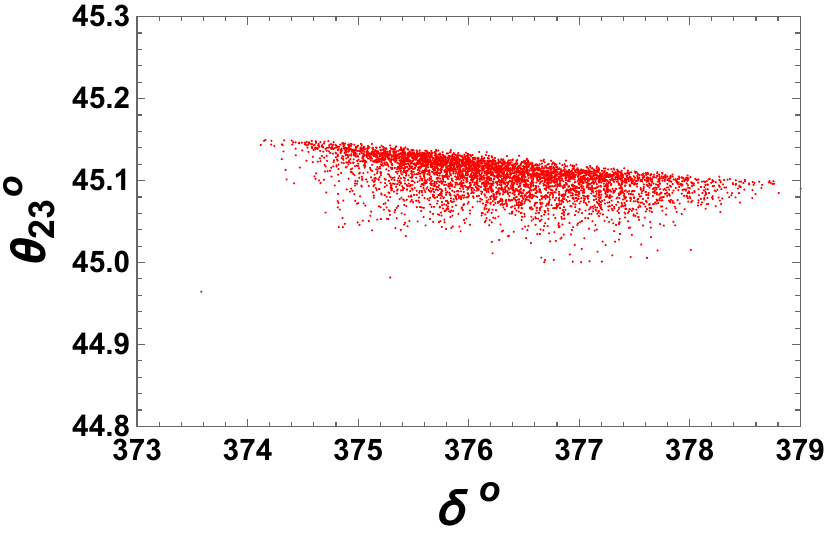}\label{fig:2(a)}} 
    \subfigure[]{\includegraphics[width=0.30\textwidth]{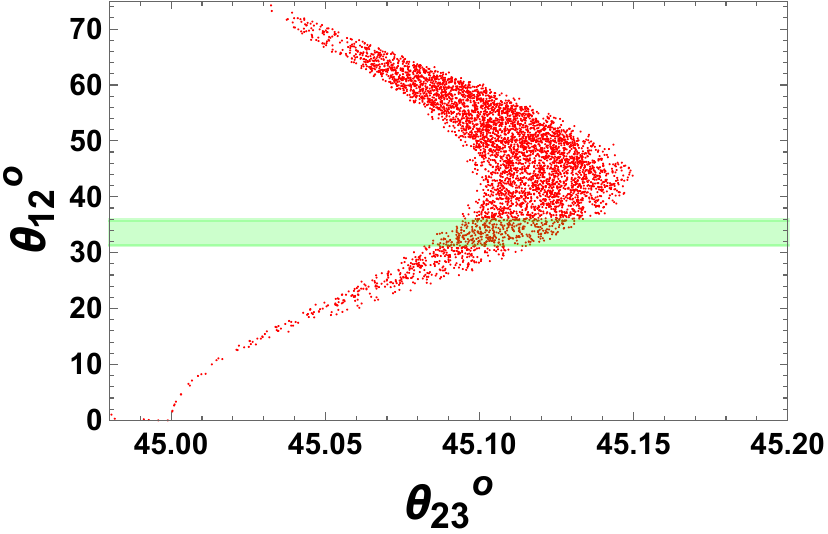}\label{fig:2(b)}} 
    \subfigure[]{\includegraphics[width=0.30\textwidth]{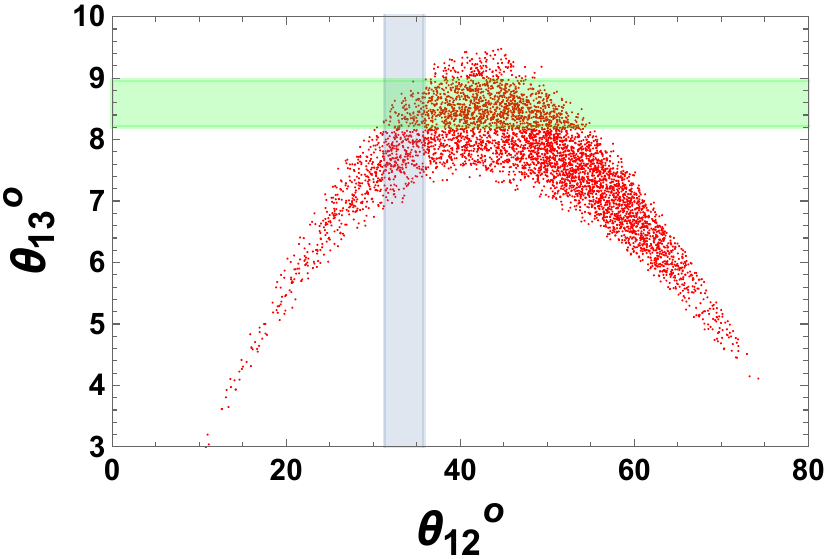}\label{fig:2(c)}}
    \caption{The correlation plots between (a) $\theta_{23}$ and $\delta$. (b) $\theta_{12}$ and $\theta_{23}$. (c)  $\theta_{13}$ and $\theta_{12}$ for normal ordering of neutrino masses with $m_3 = 0.051 \,eV$.}
\label{fig:2}
\end{figure*}

The two constraints, appearing in the mass matrix in Eq.\,(\ref{Mass matrix 1}) tells about four equations as shown below,
\begin{eqnarray}
\label{Equation 18}
Real\,[(M_{\nu})_{12}]+ Real\,[(M_{\nu})_{13}]&=& 0,\\
Im\,[(M_{\nu})_{12}]+ Im\,[(M_{\nu})_{13}]&=& 0,\\
Real\,[(M_{\nu})_{22}]- Real\,[(M_{\nu})_{33}]&=& 0,\\
Im\,[(M_{\nu})_{22}]+ Im\,[(M_{\nu})_{33}]&=& 0,
\label{Equation 21}
\end{eqnarray}
and hence, we expect the prediction of four observable parameters. As mentioned earlier, we concentrate on normal ordering and based on the oscillation data\,\citep{Gonzalez-Garcia:2021dve} and cosmological upper bound on the sum of three neutrino masses which is $0.12\,eV$ \citep{Planck:2018vyg}, we see that upper bound on $m_3$ could be approximately $0.06\,eV$, whereas the lower bound is approximately $0.05\,eV$. First, we fix the $m_3$ at $0.06\,eV$.
 Based on the conditions mentioned in Eqs.\,(\ref{Equation 18}) to\,(\ref{Equation 21}), we generate sufficiently large amount of random numbers for $\theta_{12}$, $\theta_{13}$, $\theta_{23}$ and $\delta$, and the associated  correlation plots are generated\,(see Figs.\,\ref{fig:1(a)},\,\ref{fig:1(b)}, and \ref{fig:1(c)}). It is interesting to note that the predictions of the $\theta_{ij}$'s and $\delta$ are well fitted within the $3\sigma$ bounds\,\citep{Gonzalez-Garcia:2021dve}. Moreover, we see that \emph{$\mu$-$\tau$ mixed symmetry} specifies very sharp bounds on $\theta_{23}$ and $\delta$ which are $45.087^{\circ}\leq \theta_{23}\leq 45.104^{\circ}$\,(showing the angle to lie slightly in the upper octant) and $395.038^{\circ}\leq\delta \leq 395.692^{\circ}$ respectively. In order to achieve this precision, the bounds for the mass squared differences are tuned within the $3\sigma$ ranges as per Ref.\,\citep{Gonzalez-Garcia:2021dve}, and the other input arbitrary phases are given certain numerical bounds\,(see Set 1 of Table\,\ref{table:1}). In order to see whether the similar mixing pattern could be obtainable for the lower value of $m_3$, we redo the whole exercise by fixing $m_3$ at $0.051\,eV$. The details of the numerical ranges prescribed for the input parameters are shown in Set 2 of Table\,\ref{table:1}. From, Figs.\,\ref{fig:2(a)},\,\ref{fig:2(b)}, and \ref{fig:2(c)}, we see that the proposed texture is viable for lower value of $m_3$ as well. 

To achieve the desired texture appearing in Eq.\,(\ref{Mass matrix 1}), we extend the field content of SM in terms of right-handed neutrinos\,($\nu_{l_R}$), complex scalar field doublets\,($\Phi$), a scalar singlet\,($\zeta$) and scalar triplets\,($\Delta$) in addition to the left-handed lepton doublets\,($D_{l_L}$), right-handed lepton singlets\,($l_R$) and complex Higgs doublets\,($\phi$). Further, we associate the discrete flavour symmetry $A_4$ \citep{Altarelli:2010gt, King:2006np} and define the transformation properties of the field content as shown in Table\,\ref{table:3}. In addition, we introduce the $Z_3$ symmetry in the framework in order to cut short some undesired terms in $\mathcal{L}_{Y}$ which are invariant under $A_4$. The $SU(2)_{L} \times A_{4} \times Z_{3}$  invariant Yukawa Lagrangian is framed in the $S$ basis\,\cite{Altarelli:2010gt, King:2006np} and it is presented as shown below,   
\begin{eqnarray}
- \mathcal{L}_Y &=& y_e (\bar{D}_{l_L}\phi)_1\,e_{R} + y_{\mu} (\bar{D}_{l_L}\phi)_{1'} \,\mu_{R_{1''}}+\,y_{\tau}\nonumber\\&&\,(\bar{D}_{l_L} \phi)_{1''} \tau_{R_{1'}} + y_D\,[(\bar{D}_{l_L}\Phi)_1\,\nu_{e_R} + (\bar{D}_{l_L}\nonumber\\&&\,\Phi)_{1'} \,\nu_{{\tau}_{R_{1''}}} + (\bar{D}_{l_L}\Phi)_{1''}\,\nu_{{\mu}_{R_{1'}}}]+\,\frac{1}{2}\,y_{R_1}\nonumber\\&&\,(\overline{\nu^c}_{e_R}\,\nu_{e_R})_1\,\zeta + \frac{1}{2}\,y_{R_2}\,[(\overline{\nu^c}_{\mu_{R}}\,\nu_{\tau_{R}})\,+ (\overline{\nu^c}_{\tau_{R}}\nonumber\\&&\,\nu_{\mu_{R}})]\,{\zeta} +\, y_{T_2}\,(\bar{D}_{l_L}\, D_{l_L}^c)_{3_S}\,{\Delta}_3 +\, h.c.\nonumber\\
\label{Yukawa Lagrangian}
\end{eqnarray}
The details of the multiplication rules under $A_4$ symmetry can be found in the Ref\,\cite{Altarelli:2010gt, King:2006np, King:2013eh}. We choose the vacuum expectation values (vev) in the following way, $\langle\phi\rangle_{\circ}=v_{\phi}\,(1,1,1)^{T}$\,\cite{Pramanick:2017wry} and $\langle\Phi\rangle_{\circ}= v_{\Phi}\,(1,1,1)^{T}$, $\langle\zeta\rangle_{\circ}=v_{\zeta}$ and obtain the charged lepton mass matrix, Dirac neutrino mass matrix\,($M_D$) and right-handed neutrino mass matrix\,($M_R$) as shown below,
\begin{eqnarray}
& M_{l} = v_{\phi}
 \begin{bmatrix}
 y_{e} & y_{\mu} & y_{\tau}\\
 y_{e} & \omega^2 \,y_{\mu} & \omega \,y_{\tau}\\
 y_{e} & \omega \,y_{\mu} & \omega^2 \,y_{\tau}\\
 \end{bmatrix},&
 \label{Mass matrix 2}\\
& M_{D} = y_{D}v_{\Phi}
 \begin{bmatrix}
 1 & 1 & 1\\
 1 & \omega & \omega^2 \\
 1 & \omega^2  & \omega \\
 \end{bmatrix},&
 \label{Mass matrix 3}\\
& M_{R}=
 \begin{bmatrix}
 A & 0 & 0\\
 0 & 0 & B \\
 0 & B  & 0 \\
 \end{bmatrix},&
 \label{Mass matrix 4}
\end{eqnarray}
where, $\omega= e^{i\,2\pi/3}$, $\omega^2= \omega^*$, $A=y_{R_{1}}v_{\zeta}$ and $B=y_{R_{2}}v_{\zeta}$. We extract the contribution from Type-I seesaw mechanism \,\cite{Brdar:2019iem, Mohapatra:2004zh} as $M_{T_{1}}=\,-\,M_{D}M^{-1}_{R}M^{T}_{D}$. The $\Delta$ contributes towards the Type-II seesaw\,\cite{Melfo:2011nx} and on choosing a vev, $\langle\Delta\rangle_{\circ}=v_{\Delta}(0,1,-1)^{T}$ which appears similar to that in Ref.\,\cite{Ma:2011yi}, we obtain the contribution from Type-II mechanism as shown below, 

\begin{equation}
 M_{T_2} = t
 \begin{bmatrix}
 0 &-1 & 1\\
-1 &  0 & 0 \\
  1 &  0  & 0 \\
 \end{bmatrix}, 
 \label{Mass matrix 5}
\end{equation}
where, $t = y_{T_{2}}v_{\Delta}$. The neutrino mass matrix, $M_{\nu s}$ is constructed as, $M_{\nu s} = M_{T_{1}}+ M_{T_{2}}$, in the symmetry basis. In order to shift to the flavour basis, we diagonalize $M_l$ appearing in Eq.\,(\ref{Mass matrix 2}): $M^{diag}_{l}=\,U^{\dagger}_{l_L}M_{l}U_{l_R}$, where $M^{diag}_{l}=\,\sqrt{3}\,v\,diag\,(y_e,\,y_{\mu},\,y_{\tau})$. The $U_{l_L}$ and $U_{l_R}$ can  be written as in the following,
\begin{eqnarray}
 U_{l_L} &=& \frac{1}{\sqrt{3}}
 \begin{bmatrix}
 e^{i\eta} & e^{-i\kappa} & e^{-i\xi}\\
 e^{i\eta} & \omega^2\,e^{-i\kappa}  & \omega\,e^{-i\xi} \\
 e^{i\eta} & \omega\,e^{-i\kappa}  & \omega^2\,e^{-i\xi} \\
 \end{bmatrix}, 
 \label{Mass matrix 6}\\
 U_{l_R} &=& diag( e^{i\eta},\, e^{-i\kappa},\, e^{-i\xi}).
\end{eqnarray}
The appearance of arbitrary phases, $\eta$, $\kappa$ and $\xi$ in $U_{l_L}$ and $U_{lR}$ is guided by the fact that the  choice of eigenvectors or diagonalizing matrix is not unique. When $\eta\,=\,\kappa\,=\,\xi\,=\,0$, the texture of $U_{l_L}$ reduces to a simple form similar to that in Ref.\,\cite{Babu:2002dz, Grimus:2017itg, Cabibbo:1977nk}. Keeping $\eta$ as arbitrary and on choosing $\kappa\,=\xi=\,\frac{\pi}{2}$, the final neutrino mass matrix in the flavour basis, after simplification, reduces to the form as shown below,

\begin{equation}
M_{\nu}= U^{T}_{l_L} M_{\nu_s}  U_{l_L}=
\begin{bmatrix}
 a & -b & b\\
 -b & c & d\\
 b & d & c^{*}\\
 \end{bmatrix},
\label{Reconstructed Mass matrix 1}
\end{equation}

In sum, a promising neutrino mass matrix texture, $\mu$-$\tau$ mixed symmetry is proposed which restricts the $\theta_{23}$ and $\delta$ within narrow bounds in addition to the successful prediction of other mixing angles. It is to be highlighted that the analysis is done in the light of normal ordering of neutrino masses. We have tried to obtain the said texture in the framework of $SU(2)_L \times A_4\times Z_3$, by introducing additional scalar fields and invoking Type-I and Type-II seesaw mechanisms. In the present work, we have highlighted the importance of the arbitrary phases appearing inside the $U_{l_L}$ to obtain the required texture of the neutrino mass matrix.

\section*{Acknowledgement}
One of the authors, PC thanks M Ricky Devi, Gauhati University for fruitful discussions. SR thanks L Lavoura, CFTP\,(Lisbon), IST\,(Lisbon) for important inputs. MD acknowledges support from Council of Scientific and Industrial Research\,(CSIR), Government of India through a NET Junior Research Fellowship vide grant No. 09/0059(15346)/2022-EMR-I. PC acknowledges support from the Innovation in Science Pursuit for Inspired Research (INSPIRE), Department of Science and Technology, Government of India, New Delhi vide grant No. IF190651.

	\begin{table}
\centering
\begin{tabular}{P{2cm}P{2.4cm} P{2.4cm}}
\hline 
Parameters & Set 1 & Set 2\\
\hline
$m_3/eV$ & 0.060 & 0.051\\
\hline
$\frac{\Delta\,m_{21}^2}{10^{-5}eV^2}$ & 7.27-7.45 & 6.9-8.0\\
\hline
$\frac{\Delta\,m_{31}^2}{10^{-3}eV^2}$ & 2.49-2.53 & 2.50-2.537 \\
\hline
$\phi_2/^\circ$ & 72.8-73 & 71.3-71.9\\
\hline
$\phi_3/^\circ$ & 128.9-129.1 &  276.6-277.2\\
\hline 
$\alpha/^\circ$ & 159-159.2 & 40.16-40.20\\
\hline
$\beta/^\circ$ & 142.7-142.9 & 37.81-37.86\\
\hline
\end{tabular}
\caption{ Numerical values of the inputs parameters} 
\label{table:1}
\end{table}
\vfill
\newcolumntype{P}[1]{>{\centering\arraybackslash}p{#1}}
\begin{table}[!]
\centering
\begin{tabular}{P{1.15cm}|P{0.25cm}P{1.3cm}P{1.7cm}P{0.3cm}P{0.2cm}P{0.2cm}P{0.2cm}} 
\hline
Fields & $D_{l_{L}}$ & $l_{R}$ & $\nu_{l_R}$ & $\phi$ & $\Phi$ & $\zeta$ & $\Delta$\\ 
\hline
$SU(2)_{L}$ & 2 & 1 & 1 & 2 & 2 & 1 & 3 \\
\hline
$A_{4}$ & 3 & (1,\,1{$''$},\,1{$'$}) & (1,\,1{$'$},\,1{$''$}) & 3 & 3 & 1 & 3 \\
\hline
$Z_{3}$ & 1 & ($\omega$,\,$\omega$,\,$\omega$) & ($\omega^{2}$,\,$\omega^{2}$,\,$\omega^{2}$) & $\omega^{2}$ & $\omega$ & $\omega^{2}$ & 1\\
\hline
\end{tabular}
\caption{ The transformation properties of various fields under $SU(2)_{L} \times A_{4} \times Z_{3}$.} 
\label{table:3}
\end{table}
\pagebreak

\biboptions{sort&compress}

\end{document}